\documentclass[twocolumn,prl, showpacs, amsmath,amssymb, superscriptaddress]{revtex4-1}
\usepackage{graphicx}
\usepackage{dcolumn}
\usepackage{bm}
\usepackage{color}
\usepackage[T1]{fontenc}
\usepackage[colorlinks,linkcolor=blue,citecolor=blue,urlcolor=blue]{hyperref}
\usepackage{soul}
\usepackage[normalem]{ulem}
\usepackage{color}

\begin{document}

\pdfstringdefDisableCommands{\let\sout\relax}

\title{Coherent control of the optical absorption in a plasmonic lattice coupled to a luminescent layer}

\author{Giuseppe Pirruccio}
\altaffiliation[current address: ]{Universidad Nacional Autonoma de Mexico, Mexico D.F. 01000, Mexico}
\affiliation{Center for Nanophotonics, FOM Institute AMOLF, c/o Philips Research Laboratories, High Tech Campus 4, 5656 AE Eindhoven, The Netherlands}

\author{Mohammad Ramezani}
\affiliation{Center for Nanophotonics, FOM Institute AMOLF, c/o Philips Research Laboratories, High Tech Campus 4, 5656 AE Eindhoven, The Netherlands}

\author{Said Rahimzadeh-Kalaleh Rodriguez}
\altaffiliation[current address: ]{Laboratoire de Photonique et de Nanostructures, LPN/CNRS, France}
\affiliation{Center for Nanophotonics, FOM Institute AMOLF, c/o Philips Research Laboratories, High Tech Campus 4, 5656 AE Eindhoven, The Netherlands}

\author{Jaime G\'{o}mez Rivas}
\email{J.GomezRivas@differ.nl}
\altaffiliation[current address: ]{Dutch Institute For Fundamental Energy Research, DIFFER, the Netherlands.}
\affiliation{Center for Nanophotonics, FOM Institute AMOLF, c/o Philips Research Laboratories, High Tech Campus 4, 5656 AE Eindhoven, The Netherlands}
\affiliation{COBRA Research Institute, Eindhoven University of Technology, P.O. Box 513, 5600 MB Eindhoven, The Netherlands}

\date{\today}

\begin{abstract}
We experimentally demonstrate the coherent control, i.e., phase-dependent enhancement and suppression, of the optical absorption in an array of metallic nanoantennas covered by a thin luminescent layer. The coherent control is achieved by using two collinear, counter-propagating and phase-controlled incident waves with wavelength matching the absorption spectrum of dye molecules coupled to the array. Symmetry arguments shed light on the relation between the relative phase of the incident waves and the excitation efficiency of the optical resonances of the system. This coherent control is associated with a phase-dependent distribution of the electromagnetic near-fields in the structure which enables a significant reduction of the unwanted dissipation in the metallic structures.
\end{abstract}

\maketitle

Methods for controlling optical absorption and spontaneous emission are at the heart of diverse fields of physics. In general, these methods can be classified in two types. One type of control thrives on the interplay between the excitation field and the energy levels of the emitter. Coherence and quantum interference are the essential ingredients of spontaneous emission control methods employed in atomic physics~\cite{scully96,knight98,Ficek99,wu2002,terzis2013} and, recently, in solid-state systems such as quantum dots using polychromatic incident fields~\cite{He15}. A second type of control is based on Purcell's remark: spontaneous emission is not an inherent property of the emitter, but it also depends on the environment~\cite{Purcell}. Within this paradigm structures are designed to modify the electromagnetic field intensity at the position of the emitter, thereby affecting its spontaneous emission~\cite{ladani,nitzan}. This second approach is attractive in the context of non-resonant molecular fluorescence. There, the excitation and emission frequency are different, and the coherence of the excitation is lost via rovibrational relaxation within the excited state manifold. Consequently, relaxation to the ground state via spontaneous emission can be modified by resonant structures such as optical antennas~\cite{anger06, kuhn06}, or non-resonant periodic structures such as photonic crystals~\cite{Yablo, John, Vos}. A major difference between the first and second type of absorption and spontaneous emission control concerns the role of the phase of the driving field. For atoms under resonant excitation, the phase of the driving field represents an important degree of freedom for controlling spontaneous emission~\cite{knight98}. In contrast, non-resonant molecular fluorescence enhancements based on designed electromagnetic environments are widely regarded as phase-insensitive.

In this manuscript we demonstrate the coherent control, i.e., phase-dependent enhancement and suppression, of the absorption and, consequently, of the non-resonant photoluminescence (PL) emission intensity from an ensemble of molecules. To achieve this, we couple an ensemble of randomly dispersed dye molecules in a polymer matrix to a periodic array of metallic nanostructures. Coherent control in plasmonic systems has been used in pioneering works of nanoscopy~\cite{Aeschlimann2011,brixner2013}, and to achieve  femtosecond and nanoscale control over electromagnetic hotspots~\cite{Aeschlimann2010,Aeschlimann2012,hulst2010}. Our array displays localized and collective electromagnetic modes weakly coupled to the molecules at the emission and absorption frequencies, respectively. Previous works have shown that the emission spectrum from similar systems can be designed by resonant processes at the emission frequencies~\cite{Vecchi09, Giannini10, Rodriguez12APL, lozano13}, while the intensity can be enhanced by processes at the absorption frequencies under single-wave illumination~\cite{lozano14}. Here, we combine these approaches by driving the coupled array-emitter system with two coherent, collinear and counter-propagating laser waves whose relative phase is controlled. The dependence of the resonances of the array on the relative phase of the driving fields allows us to demonstrate resonantly enhanced phase-dependent absorption and emission intensity in a coupled resonator-emitter system. The phase-sensitive electromagnetic field enhancements at the position of the emitters and of the metallic nanostructures are analyzed through full-wave simulations. Our central finding is that the ratio of the absorption by the molecules to the absorption by the metallic structures exhibits a maximum at a particular relative phase of the driving fields. Our results elucidate a new way to mitigate losses in plasmonic systems with respect to the usual approach employing gain media~\cite{campione11,pustovit15,deluca11,deluca12,infusino14}.

We have investigated the optical response of a $3\times3$ mm$^2$ square array of aluminum nanopyramids fabricated onto a silica substrate using substrate conformal imprint lithography~\cite{marcthesis}. The lattice constant is 340 nm, the pyramids have a height of 150 nm, are 100 nm wide at the base and 80 nm wide at the top. We spin-coated on top of the array a 200 nm-thick layer of polystyrene doped at 1\% weight concentration with F305 Lumogen$^\copyright$ dye. The layer has an internal quantum efficiency of 90\%. The absorption spectrum, exhibiting a maximum at 574 nm, is shown in the Supplemental Information (SI)~\cite{supplemental}.

\begin{figure}
\includegraphics[width=0.5\textwidth]{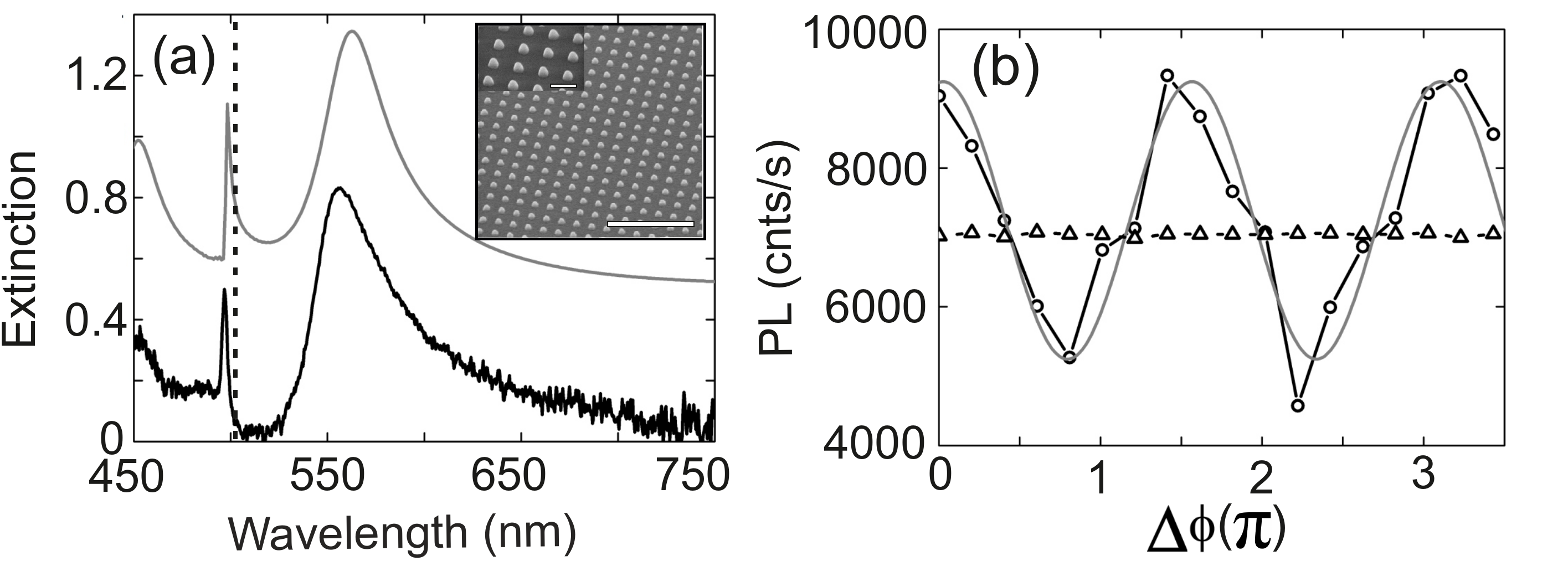}
\caption{(a) Measured (black) and simulated (grey) normal incidence extinction of the array of aluminum nanopyramids. For clarity, the simulation is vertically displaced by 0.5. The dashed line corresponds to the degenerate ($\pm$1, 0), (0, $\pm$1) RAs. Inset: SEM image of the array. The scales of the inset are 2$\mu$m (large image) and 300 nm (zoom). (b) Connected open circles: measured modulation of the maximum of the PL as a function of the phase difference between the two incident waves. Connected open triangles: PL of the dye layer resulting from the incoherent sum of the two incident waves. Grey continuous line: guide to the eye representing a cosine square function.}\label{fig1}
\end{figure}

The inset of Fig.~\ref{fig1}(a) shows scanning electron micrographs of the array prior to the deposition of the dye layer. Figure~\ref{fig1}(a) shows with black line the normal incidence extinction spectrum given by $1-T_0$, with $T_0$ the zeroth order transmittance. The grey line indicates simulations results, vertically shifted for clarity, for the same structure obtained with the Finite Difference in Time Domain method. We use periodic boundary conditions and values of the Al permittivity from Ref.~\cite{Palik}. The refractive index of the glass substrate is constant at $n_s = 1.46$, while the complex refractive index of the dye-doped polymer layer was obtained from ellipsometry measurements. The two peaks observed in experiments and simulations at wavelengths around 560 nm and 497 nm correspond to photonic-plasmonic resonances in the particle array. The peak at $\lambda$ = 560 nm is associated to localized surface plasmon resonances (LSPRs) in individual metallic nanostructures and we call this wavelength $\lambda_{LSPR}$. The peak at $\lambda$ = 497 nm corresponds to collective resonances termed as surface lattice resonances (SLRs)~\cite{said14} and this wavelength is called $\lambda_{SLR}$. The origin of the SLRs is the radiative coupling between LSPRs and the degenerate ($\pm$1, 0), (0, $\pm$1) Rayleigh Anomalies (RAs), i.e., the diffracted orders radiating grazing to the plane of the array~\cite{zou,hicks,chu,auguie,kravets,giannini,teperik,rodriguez11,alu13,campione14}. These RAs are marked with dashed line in Fig.~\ref{fig1}(a). The spatially-resolved near-field at these two resonances is shown in Fig.~S3 in the SI~\cite{supplemental}.

For the coherent control measurements we used a Mach-Zender interferometer comprising two collinear and counter-propagating continuous waves (a control and a signal) from an Ar-Kr laser emitting at the wavelength $\lambda_{SLR}$, illuminating the sample at normal incidence (see SI for the scheme of the set-up~\cite{supplemental}). We ensured that the two incident waves have equal intensity. The control wave propagating from the air-side is phase-delayed with respect to the signal wave propagating from the substrate-side. The phase difference between the two waves is controlled by changing the optical path length of the control wave with a computer-controlled piezo mirror. A similar approach was used to demonstrate time-reversed lasing in thick slabs~\cite{Chong10,Wan11} and coherent absorption in thin layers~\cite{Zhang12,pirruccio12,pirruccio13,Fan14}. However, the applicability of these techniques to the realm of coherent absorption in plasmonic systems and spontaneous emission control has hitherto remained unexplored. The extension is far from trivial because the far-field spectrum of metallic nanostructures can differ from their near-field~\cite{Greffet,Bryant,Ross,Gallinet,hentschel,stockman}, and emitters are sensitive to the latter~\cite{frimmer,RodriguezPRL12}. Moreover, in our experiments the emitters and the resonators are spatially separated, and the presence of strong field gradients can modify the absorption in the emitters with respect to the absorption in the metal.

The emission was measured at 14 degrees from the normal to the sample. At this angle the emission intensity is maximum due to the coupling of this emission to the (-1,0) SLR of the array. The measurements of the emission spectra at different angles are shown in Fig. S4(a) of the SI~\cite{supplemental}. The emitted spectrum and intensity by the dye layer depend on processes taking place at emission and absorption wavelengths. Here we focus on the latter. In Fig.~\ref{fig1}(b) we plot the peak PL intensity as a function of the phase difference between the incident waves with connected open circles. Since molecular fluorescence is an incoherent process and we are acting only at the absorption wavelength, no change in the spectral shape was observed as a function of $\Delta\phi$. Due to the interference nature of this phenomenon we expect the PL intensity to follow a cosine square function. In Fig.~\ref{fig1}(b) the grey continuous line is a guide to the eye representing a cosine square function with period slightly smaller than 2$\pi$. The deviation from 2$\pi$ is likely due to a small misalignment in the interferometer that may result in a difference between nominal read-outs of the piezo actuator and the real path difference introduced. The connected open triangles in Fig.~\ref{fig1}(b) correspond to the PL resulting from the incoherent sum of the two incident waves illuminating the sample either from the air-side or from the substrate-side.

The PL is strongly modulated by the absorption efficiency of the dye at different pump phases. This, in turn, is due to the electric field distribution in the system, which is related to the excitation efficiency of the SLR. While the single wave time-integrated dissipated power is phase-insensitive, the presence of two waves dramatically changes this situation. This phenomenon is explained in detail in what follows.

We have performed FDTD simulations to illustrate the interference mechanism that determines the absorption in the structure and gives rise to the measured PL modulation. Two separate simulations have been done with a single plane wave impinging either from the air-side or from the substrate-side. The absorption of the dye-doped polymer layer was included via the complex refractive index $n_l=1.59+i0.003$ at $\lambda_{SLR}$. The complex electromagnetic field components were calculated as a function of the relative phase between the two waves, $\Delta\phi$, using the superposition principle for the fields. In absence of the nanopyramids, the two counter-propagating incident waves form a quasi-standing wave for the wavelengths $\lambda_{LSPR}$ and $\lambda_{SLR}$, i.e., the shift in the position of its node and antinode along z is negligible compared to the height of the nanopyramid~\cite{said14}. We define $\Delta\phi$ = 0 when this quasi-standing wave has a node approximately at a height that corresponds to the mid-height of the nanopyramids. Correspondingly, for $\Delta\phi = \pi$ the total field has an antinode at this height.

\begin{figure}
\includegraphics[width=0.5\textwidth]{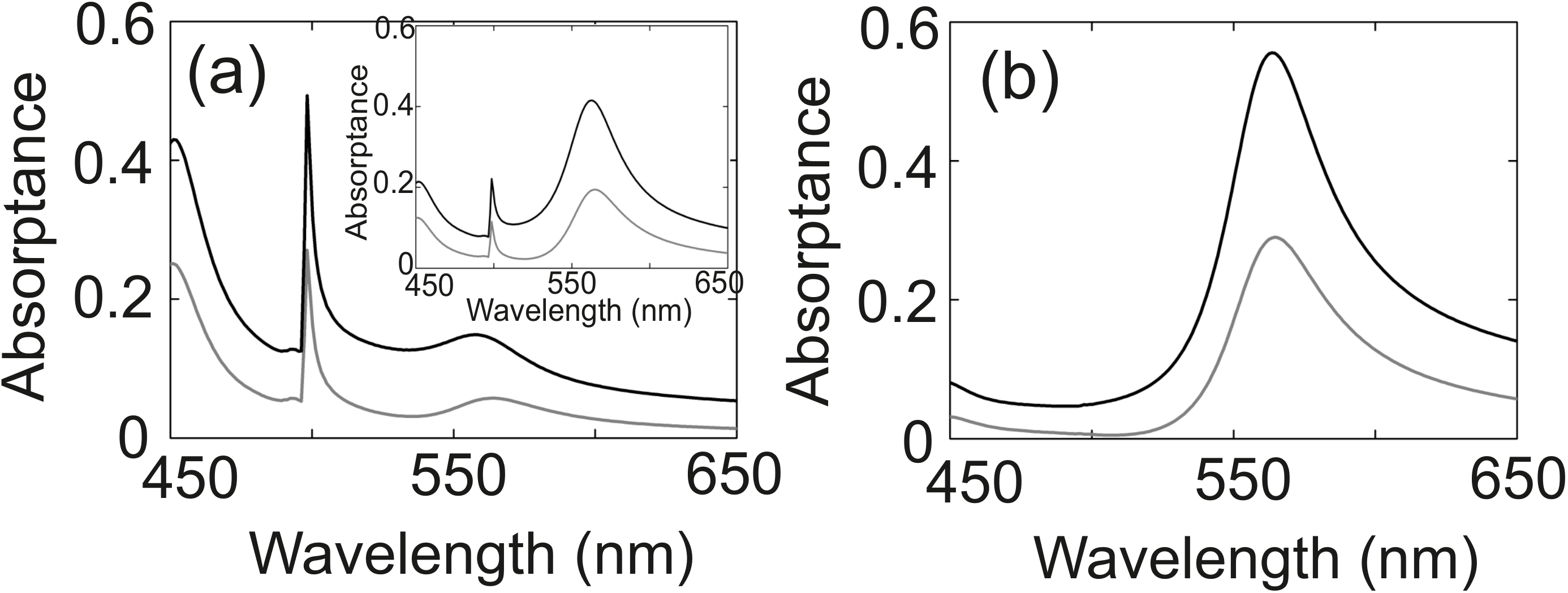}
\caption{Calculated absorptance in the dye layer (black curve) and in the metal nanopyramids (grey curve) as a function of the incident wavelength for a phase difference between the two driving fields of (a) $\Delta\phi$ = 0, (b) $\pi$. Inset: calculated absorptance for single wave illumination.}\label{fig2}
\end{figure}

The absorptance of the system as a function of incident wavelength, $\lambda$, and phase difference between the two incident waves, $\Delta\phi$, is given by $A(\lambda,\Delta\phi) = \frac{\pi c \epsilon_0}{\lambda}\int_V{\epsilon(\lambda)''|E_{tot}(\lambda,\Delta\phi)|^2dV}$, with $\epsilon_0$ the vacuum permittivity, $c$ the speed of light, $\epsilon''$ the imaginary component of the permittivity of the dye layer or of the metallic nanopyramids and $V$ the integration volume. $|E_{tot}(\lambda,\Delta\phi)|^2$ = $|E_{1}(\lambda,\phi_1)+E_{2}(\lambda,\phi_2)|^2$ is the intensity of the total electric field normalized by the incident field and $|E_{1,2}(\lambda,\phi_{1,2})|^2$ are the intensities of the two incident waves. The inset in Fig.~\ref{fig2}(a) shows the absorptance integrated over the volume of the nanopyramids (grey curve) and of the dye layer (black curve) under single plane wave illumination impinging from the air-side. This phase-independent absorptance, calculated with $E_{2}(\lambda,\phi_2)=0$, is resonantly enhanced in the metal and in the dye layer at the wavelengths corresponding to the SLR and LSPR.

In Figs.~\ref{fig2}(a) and (b) the absorptance in the dye layer and in the nanopyramids is represented for $\Delta\phi$ = 0 and $\pi$, respectively. For $\Delta\phi$ = 0 we observe that the absorptance at the SLR wavelength, i.e., at 497 nm, is enhanced both in the metal and in the dye while the absorptance at the LSPR wavelength, i.e., at 560, nm is reduced. The opposite behavior is observed for $\Delta\phi = \pi$, where the SLR absorption is fully suppressed, while that of the LSPR is significantly enhanced. These results are in agreement with the phase-modulated emission previously presented, where the emission correlates with the modulated absorption in the dye layer at the wavelength of the pump laser, i.e., $\lambda$ = 497 nm.

\begin{figure}
\includegraphics[width=0.5\textwidth]{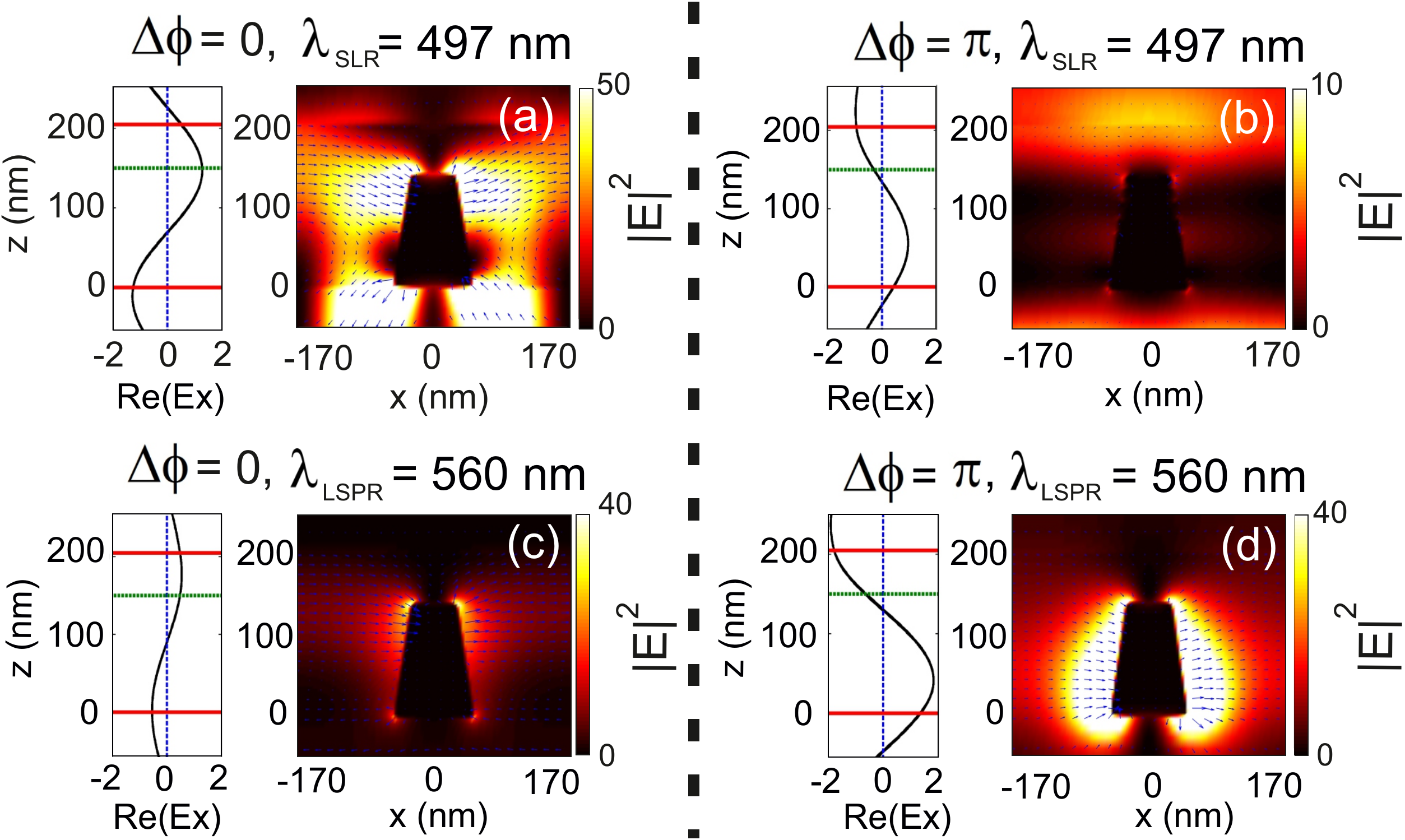}
\caption{Color plots: spatial distribution of the normalized intensity of the total electric when the system is illuminated at normal incidence with two coherent, collinear and counter-propagating waves. The field is plotted in the plane y = 0 crossing the nanopyramid along its center. Arrows: real part of the y-z component of the total electric field. (a,b) are calculated for $\lambda_{SLR}$ while (c,d) are calculated for $\lambda_{LSPR}$. (a,c) are obtained for $\Delta\phi$ = 0, while (b,d) are obtained for $\Delta\phi = \pi$. Next to each plot we show the calculated real part of the amplitude of the total incident electric field in the dye layer without the array. The horizontal red lines indicate the thickness of the dye layer, and the green dashed line represents the height position of the top of the nanopyramid. The dashed blue line indicates $Re(E_x) = 0$.}\label{fig3}
\end{figure}

Next we elucidate the coherent control over the excitation efficiency of the two resonances by analyzing the near-fields. In Figs.~\ref{fig3}(a-d) we plot in color the electric field enhancement at the plane defined by the incident polarization and wave vector. The arrows represent the real part of the $x-z$ electric field components. The incident wavelength is $\lambda_{SLR}$ in Figs.~\ref{fig3}(a,b), and $\lambda_{LSPR}$ in Figs.~\ref{fig3}(c,d). Figures~\ref{fig3}(a,c) correspond to $\Delta\phi$ = 0, while Figs.~\ref{fig3}(b,d) correspond to $\Delta\phi = \pi$ (the color scale changes from figure to figure). The fields in Figs.~\ref{fig3}(a,d), calculated for $\Delta\phi = 0$ and $\Delta\phi=\pi$, respectively, qualitatively resemble those for the single wave illumination (see Fig.~S2 in SI~\cite{supplemental}).

By comparing Figs.~\ref{fig3}(a,b) with Figs.~\ref{fig3}(c,d) we notice that there is an opposite response of the system to $\Delta \phi$ at $\lambda_{SLR}$ and $\lambda_{LSPR}$. To explain this different response we compare the symmetry of the field distribution when the two resonances are excited with the symmetry of the driving field in absence of the array. To facilitate this comparison, the real part of the $x$-component of this driving field is plotted next to each color plot in Fig.~\ref{fig3} with a continuous black curve. For both wavelengths the driving field is antisymmetric for $\Delta\phi$ = 0 and symmetric for $\Delta\phi = \pi$, both with respect to the $xy$ plane crossing the nanopyramid approximately at its mid-height. The arrows plotted in Figs.~\ref{fig3}(a,b) show a quadrupolar field distribution in the nanopyramid for $\lambda_{SLR}$, which is antisymmetric with respect to the plane $z = 40$ nm. In contrast, the field is dipolar and symmetric for $\lambda_{LSPR}$. Therefore, the SLR is efficiently excited only when the total driving field is antisymmetric and it is suppressed when the driving field is symmetric. The opposite occurs for the LSPR. We conclude that by controlling the symmetry of the field distribution via the relative phases of the driving fields, it is possible to control the efficiency of the excitation of the two resonances. The growing interest towards coherent control of resonances in nanoparticles is demonstrated by a recent work~\cite{das15}. Consequently, the absorptance shown in Fig.~\ref{fig2} depends pronouncedly on the symmetry match between driving field and optical resonances. The origin of this effect is on the height of the nanopyramids and the retardation of the scattered field with respect to the incident field (see SI~\cite{supplemental}). The field extends throughout the unit cell when both resonances are efficiently excited, but it is more confined to the nanopyramid at the LSPR wavelength. For both resonances, the regions of high and low electric field intensity interchange when passing from $\Delta\phi$ = 0 to $\Delta\phi = \pi$.

\begin{figure}
\includegraphics[width=0.5\textwidth]{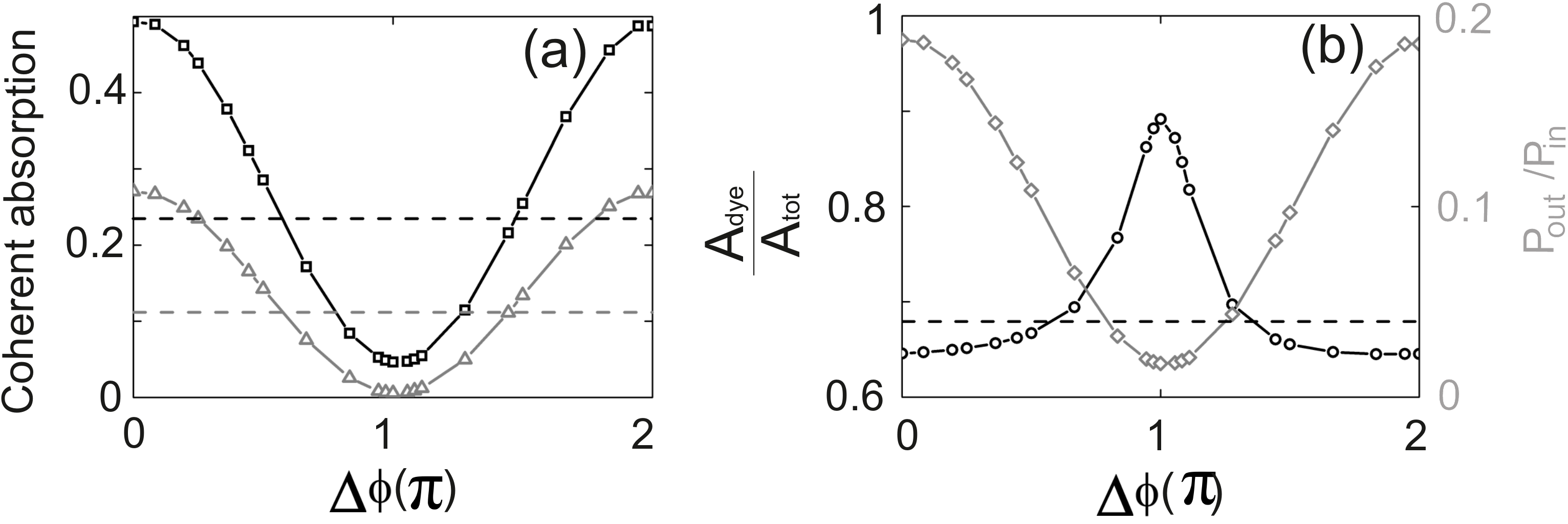}
\caption{(a) Modulation of the absorptance in the dye layer (open squares and black curve) and in the metallic array (open triangles and grey curve) for $\lambda_{SLR}$ as a function of $\Delta\phi$. Dashed lines: incoherent absorption in the dye layer (black line) and in the metallic array (grey line) (b) Open circles and black curve: ratio of the absorption in the luminescent layer by the total absorption, as a function of $\Delta\phi$. Dashed horizontal line: condition of single wave illumination. Open diamonds and grey curve: $P_{out}/P_{in}$ as a function of $\Delta\phi$.}\label{fig4}
\end{figure}

The change in the spatial distribution of the near field intensity in Fig.~\ref{fig3} implies a change in the relative absorption between nanopyramids and dye layer. To assess this phase-dependent change we calculate the absorptance in the volume occupied by the emitters, i.e. the dye layer, $A_{dye}$, and the absorptance in the metallic nanopyramids, $A_{metal}$. In Fig.~\ref{fig4}(a) we plot $A_{dye}$ (open squares and continuous black curve) and $A_{metal}$ (open triangles and continuous grey curve) as a function of $\Delta\phi$ for the SLR wavelength. The dashed black and grey lines correspond to the absorption in the dye layer and in the metallic array, respectively, for single wave illumination. Figure~\ref{fig4}(a) shows that $A_{dye}$  and $A_{metal}$ are both  a cosine square function of $\Delta\phi$. The theoretical modulation of the absorption qualitatively agrees with the experimental modulation of the emission in Fig.~\ref{fig1}(b), but the latter is roughly three times lower. This is likely due to the fact that the experimental pump wavelength coincides with the maximum in extinction at the SLR, but the maximum in absorption is slightly shifted by 2 nm with respect to the extinction. A small surface roughness of the dye layer ($\approx\pm 20$ nm) could also modify the light in-coupling and out-coupling~\cite{Pirruccio15}.

In Fig.~\ref{fig4}(b) we plot the absorption ratio $\frac{A_{dye}}{A_{tot}}=\frac{A_{dye}}{A_{metal}+A_{dye}}$ as function of $\Delta\phi$. The dashed line refers to the single wave illumination. The maximum absorption ratio that can be theoretically achieved in the investigated sample reaches a value as high as 90\%, significantly improving the 68\% of a single wave. The improved FoM is the consequence of the reduced pump enhancement by the nanostructures, which also leads to a net reduction in the absorption of the dye layer as shown in Fig.~\ref{fig4}(a). The improved absorption ratio results from a strongly reduced absorption in the metal and a significant, albeit reduced, absorption in the dye layer. This is associated with a redistribution of the near field intensity in the structure which, for the $\Delta\phi$ corresponding to the the peak of the absorption ratio, overlaps better with the dye layer than with the metallic nanopyramids. In Fig.~\ref{fig4}(b) we plot the ratio of the total output power to the incident one with grey curve and open diamonds (see SI for details on the calculation~\cite{supplemental}). This ratio is reduced for $\Delta\phi=\pi$ as the absorption in the dye is reduced with respect to the single wave configuration. However, the absorption and, consequently, the light conversion efficiency can be optimized by changing the excitation angle~\cite{guo}, or by increasing the dye concentration. Since the coherent control of the absorption ratio is only achieved at absorption wavelengths, any process occurring at the emission wavelengths is preserved. This emission can still be enhanced at the corresponding wavelengths, (see Fig. S4~\cite{supplemental}).~\cite{Vecchi09}.

In conclusion, we have experimentally demonstrated the coherent control of absorption and, as a consequence, the modulation of light emission in an ensemble of dye molecules coupled to an array of aluminum nanopyramids illuminated with two coherent, collinear, counter-propagating and phase-controlled waves. Photoluminescence intensity measurements show a strong dependence on the relative phase between these waves, in qualitative agreement with the enhanced absorption in the dye layer obtained with numerical simulations. The coherent control of absorption is mediated by the enhancement and suppression of hybrid plasmonic-photonic resonances in arrays of metallic nanopyramids, as a consequence of the match between the symmetries of these modes and of the driving field. By controlling the phase-dependent optical absorption of the system we can achieve a significant reduction of losses in plasmonic structures. As it relies on symmetry, our approach is general with respect to the emitters used and to the resonant structure that they are coupled to.

This work was supported by the Netherlands Foundation Fundamenteel Onderzoek der Materie (FOM) and the Nederlandse Organisatie voor Wetenschappelijk Onderzoek (NWO), and is part of an industrial partnership program between Philips and FOM.




\newpage
\newpage

\newpage

\newpage

\newpage

\end{document}